\documentclass[aps,prm,reprint, superscriptaddress]{revtex4-2}
\usepackage{blindtext}
\usepackage{graphicx}% Include figure files
\usepackage{xcolor}
\usepackage{dcolumn}% Align table columns on decimal point
\usepackage{bm}% bold math
\usepackage[utf8]{inputenc}
\usepackage{natbib}

\begin{document}
	
	\preprint{APS/123-QED}
	
	\title{Intercalation induced quasi-freestanding layer in TiSe$_2$}
	
	\author{Turgut Yilmaz}
	\affiliation{Department of Physics, Xiamen University Malaysia, Sepang 43900, Malaysia}
	\affiliation{Department of Physics, University of Connecticut, Storrs, CT 06269, USA}
	\email{trgt2112@gmail.com}
	
	\author{Yi Sheng Ng}
	\affiliation{Department of New Energy Science and Engineering, Xiamen University Malaysia, Sepang 43900, Malaysia}
	
	\author{Anil Rajapitamahuni}
	\affiliation{National Synchrotron Light Source II, Brookhaven National Lab, Upton, New York 11973, USA}
	
	\author{Asish Kundu}
	\affiliation{National Synchrotron Light Source II, Brookhaven National Lab, Upton, New York 11973, USA}
	
	\author{ Hui-Qiong Wang}
	\affiliation{Department of Physics, Xiamen University, Xiamen 361005, P. R. China}
	\affiliation{Department of Physics, Xiamen University Malaysia, Sepang 43900, Malaysia}
	
	\author{Jin-Cheng Zheng}
	\affiliation{Department of Physics, Xiamen University, Xiamen 361005, P. R. China}
	\affiliation{Department of Physics, Xiamen University Malaysia, Sepang 43900, Malaysia}
	
	\author{Elio Vescovo}
	\affiliation{National Synchrotron Light Source II, Brookhaven National Lab, Upton, New York 11973, USA}

	\date{\today}% It is always \today, today,
	%  but any date may be explicitly specified

	\begin{abstract}
		
		Angle-resolved photoemission spectroscopy is employed to study the electronic structure of bulk TiSe$_2$ before and after doping with potassium impurities. A splitting in the conduction band into two branches is observed after room-temperature deposition. The splitting energy increases to approximately 130 meV when the sample is cooled to 40 K. One branch exhibits a non-dispersive two-dimensional feature, while other one shows the characteristics of three dimensional bulk band dispersion. Core level spectroscopy suggests that the K impurities predominantly occupy the intercalated sites within the van der Waals gap. The results indicate the formation of a quasi-freestanding TiSe$_2$ layer. Additionally, doping completely suppresses the periodic lattice distortion in the surface region. These findings are further supported by density functional theory calculations, which compare the band structure of monolayer and bulk TiSe$_2$ with experimental data. Thus, the dimensional and intrinsic electronic properties of 1T-TiSe$_2$ can be controlled through the intercalation procedure used in this work.
		
	\end{abstract}
	
	% Classification Scheme.
	%\keywords{Suggested keywords}%Use showkeys class option if keyword
	%display desired
	\maketitle
	
	%\tableofcontents

	Layered transition metal dichalcogenides (TMDCs) are prime examples of quasi-two-dimensional materials \cite{choi2017recent, manzeli20172d}. Each layer consists of a transition metal sheet sandwiched between two identical chalcogen layers, with the layers held together by weak van der Waals (vdW) interactions. This low-dimensional nature, combined with a rich variety of physical phenomena, has led to extensive studies on their transport, magnetic, and electronic properties. One of the most notable characteristics of TMDCs is the periodic lattice distortion (PLD), which is often associated with the charge density wave (CDW) phase \cite{neto2001charge, rossnagel2011origin, wilson1975charge}.
	
	TiSe$_2$ is a particularly intriguing material whose electronic structure has been studied both theoretically and experimentally for decades \cite{wilson1977concerning, fang1997bulk, negishi2006photoemission, rasch20081, stoffel1985experimental}. It undergoes a structural transition at 202 K, followed by a periodic lattice distortion with the formation of a 2$a$ $\times$ 2$b$ $\times$ 2$c$ supercell \cite{stoffel1985experimental,di1976electronic}. The surface electronic structure is characterized by a conduction band formed by Ti 3d atomic orbitals at the $L$-points of the hexagonal Brillouin zone, while the valence bands, which are primarily composed of Se 4p atomic orbitals, are located at the $\Gamma$-point. The lattice distortion can be monitored in angle-resolved photoemission spectroscopy (ARPES) experiments with the observation of folded valence bands (FVBs) onto the hexagonal Brillouin zone corners.

	To date, the origin of the structural distortion in TiSe$_2$ remains a topic of debate. Key scenarios include the Jahn-Teller effect \cite{hughes1977structural}, the excitonic phase \cite{pillo2000photoemission, cercellier2007evidence, monney2010probing}, and their combination \cite{van2010exciton}. However, recent findings have shown that the bulk band gap vanishes in the distorted phase, which challenges the excitonic phase scenario \cite{yilmaz2023gapless}. Additionally, several studies have focused on the electronic structure of TiSe$_2$ under the surface deposition of K impurities in the low-temperature phase to explore the origin of PLD \cite{jaouen2023carrier,jeong2024dichotomy}. Two-dimensional electron gas and band-selective electron doping are reported.
	
	Building on these insights, the broader context of alkali metal intercalation emerges as a powerful approach for tailoring the properties of TMDCs through various mechanisms, such as charge doping \cite{gong2018spatially}, expansion of the c-axis lattice parameter \cite{zhu2016tuning, wang2018monolayer}, and orbital hybridization \cite{zhao2020engineering}. These modifications enable the tuning or enhancement of a broad spectrum of properties, including electrical conductivity \cite{kappera2014metallic}, magnetic ordering \cite{zhao2020engineering, kappera2014phase}, and thermoelectric behavior \cite{nair2020electrical}. In particular, K intercalation has recently attracted significant attention for its potential in energy storage applications, especially due to its efficiency in ion battery systems \cite{zhang2019two, chen2020transition, zhou2022molybdenum}. Therefore, alkali metal deposition on the surface of TMDCs is a long-studied topic in photoemission studies \cite{zhang2020intercalation,jung2016intercalation,friend1987electronic,starnberg2000photoemission}. Earlier research on TiSe$_2$ and other TMDCs has provided evidence of intercalated impurities when deposition occurs at room temperature. This method can effectively control the transport properties and dimensionality of the materials. In particular, a quasi-freestanding layer can be achieved if the spacing between layers is sufficiently increased.  However, very few materials exhibit a strong layer splitting as a result of intercalated impurities \cite{eknapakul2014electronic, rybkina2021quasi, nakata2019dimensionality}.

	\begin{figure*}[t]
		\centering
		\includegraphics[width=15cm,height=8.416cm]{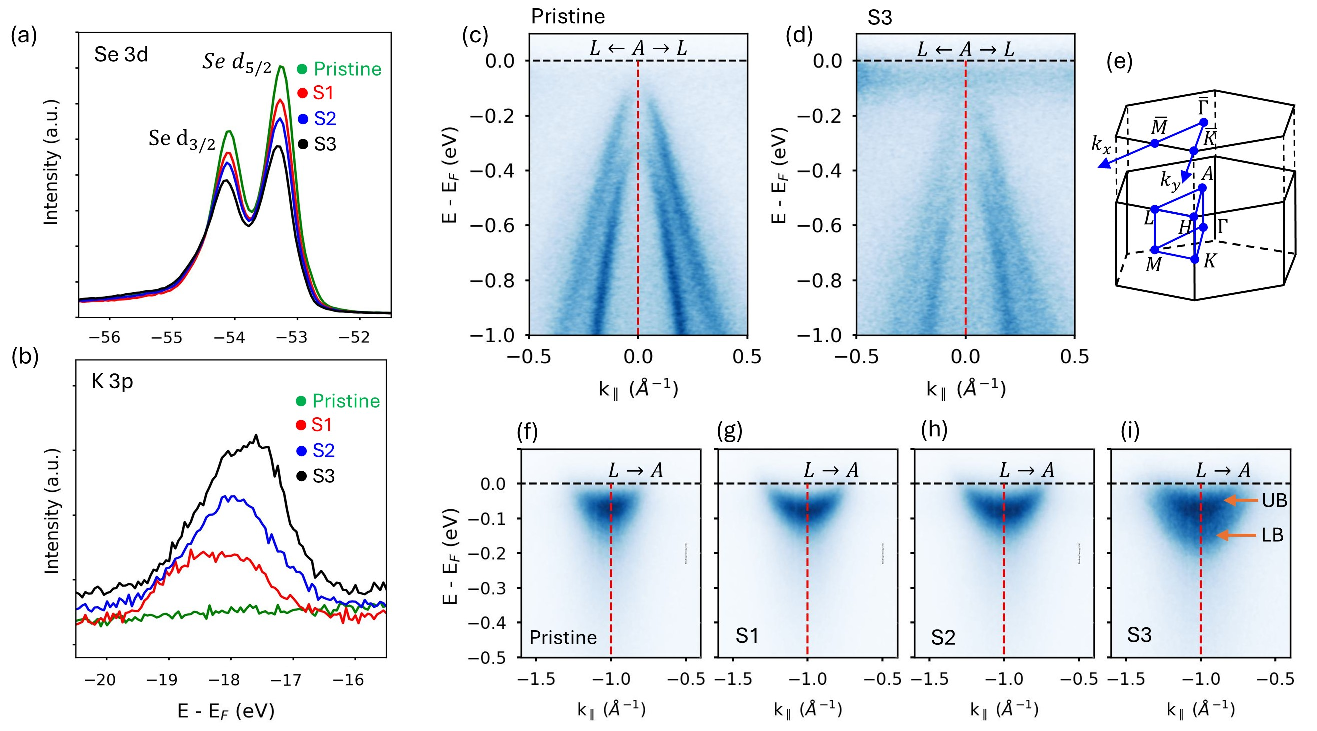}
		\caption{
			(a-b) Se 3d and K 3p core levels as a function of K-deposition level, respectively. (c-d) ARPES spectra along the $L$-$A$-$L$ direction of the Brillouin zone taken from pristine and K-doped samples, respectively. (e) 2D and 3D representation of the hexagonal Brillouin zone of TiSe$_2$. (g-j) ARPES spectra as a function of increasing K-doping level. All spectra are taken at room temperature with 119 eV linear horizontal polarized lights.
		}
	\end{figure*}

	In this study, we applied micro angle-resolved photoemission spectroscopy ($\mu$-ARPES) to investigate the electronic structure of TiSe$_2$ following the room-temperature deposition of potassium (K) impurities on the sample surface. The deposition results in a weak band splitting of the conduction band into two branches, with the energy splitting notably increases when the sample is cooled to 40 K. Photon energy-dependent ARPES data reveal that one branch exhibits two-dimensional (2D) characteristics, while the other shows bulk behavior with three-dimensional (3D) electronic dispersion. Core level spectroscopy at room temperature suggests that K impurities penetrate the sample and occupy the van der Waals (vdW) gaps. Cooling the sample increases the number of K impurities in the top vdW gap relative to the sample surface This expected to enhance the spacing between the top layer and the rest of the material and lead to the formation of a quasi-freestanding layer of TiSe$_2$. Furthermore, the PLD is completely suppressed at low temperatures. These experimental results are supported by density functional theory calculations of the band structure for both monolayer and bulk TiSe$_2$. The suppression of PLD in reduced dimensions may facilitate the realization of low-temperature phenomena such as ferromagnetic and superconducting orders, which compete with lattice distortion.

	Commercial single-crystal 1T-TiSe$_2$ samples are obtained from 2dsemiconductors. $\mu$-ARPES and core level experiments were conducted at 21-ID-1 (ESM) beamline of NSLS-II using a DA30 Scienta electron spectrometer. The beam spot size was maintained at approximately 5 $\mu$$^2$ \cite{rajapitamahuni2024electron}. K impurities were deposited at room temperature from an outgassed source. Due to the rapid intercalation of the alkali metal atoms and the variation in the sticking coefficient at room temperature, precise calibration of the deposition rate was not achievable. But three depositions were performed at evaporator currents of 4.5 A, 4.7 A, and 4.9 A. Each deposition lasted 10 minutes, and the samples are referred to as S1, S2, and S3 throughout the article. To provide a more accurate estimate of the deposition rate, we repeat K doping at 40 K sample temperature, as K atoms are expected to remain localized on the sample surface at low temperatures. We then record the core levels (see Figure S1 in Ref. \cite{SI}). The K coverage was calculated based on the total area of the spectral peaks and the photoionization cross-section. This yields 0.2, 0.34, and 0.7 monolayers of K atoms after 10 minutes of deposition at evaporator currents of 4.5 A, 4.7 A, and 4.9 A, respectively.

	Band structure calculations were performed with the Quantum Espresso (QE) package, based on density functional theory within the Perdew–Burke–Ernzerhof (PBE) parameterization of the exchange-correlation functional \cite{giannozzi2009quantum,giannozzi2017advanced}. High symmetry points of the bulk Brillouin zone for un-distorted phase are used through the article.  The kinetic energy cutoff was set to 350 eV for all the calculations and 13$\times$13$\times$13 k-mash was utilized for both bulk and monolayer samples. A vacuum level of 15 \r{A} is added to model the band structure of TiSe$_2$ monolayer.
	
	A K-intercalated TiSe$_2$ system is modeled using a 2$\times$2$\times$1 supercell with a K atoms positioned on top of a Ti atoms, which was recently shown to be energetically favorable site\cite{jaouen2023carrier}. The supercell was fully relaxed until the residual force on each ion was less than 0.01 \r{A}$^{-1}$. The electronic self-consistent iteration converged to a precision of 10$^{-6}$ eV in total energy for the K-intercalated system, and a 4$\times$4$\times$4 k-mesh was used to study the band structure. Band structure of the supercell were unfolded using unfold.x code \cite{popescu2012extracting}.

	\begin{figure*}[t]
		\centering
		\includegraphics[width=17cm,height=6.638cm]{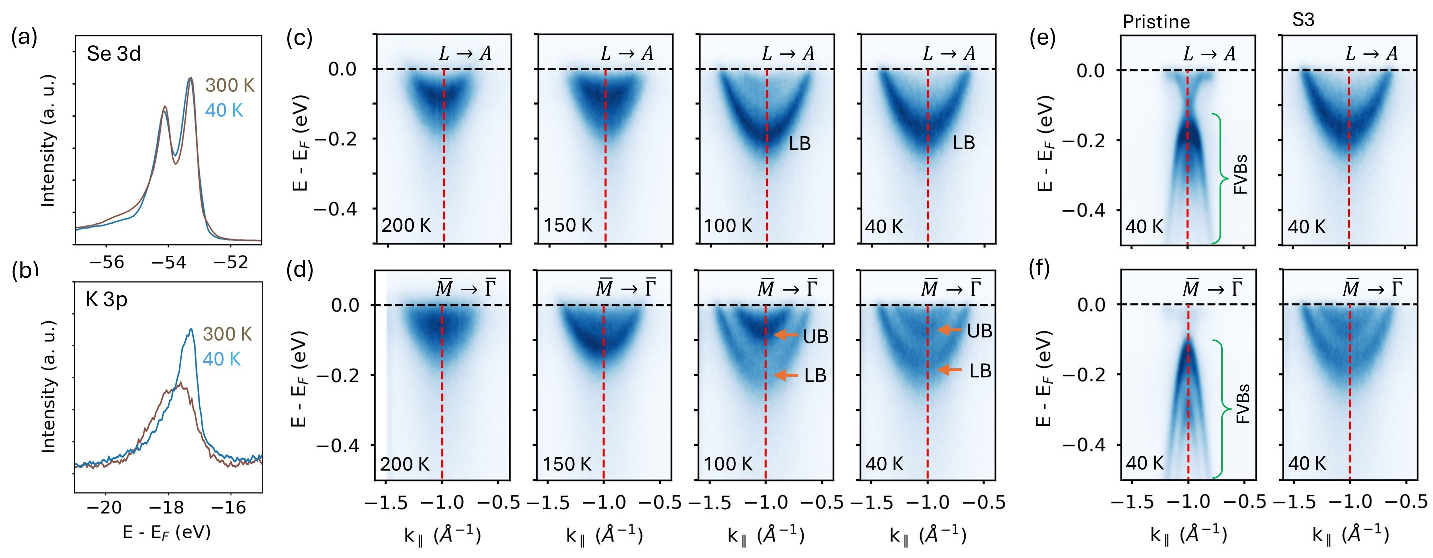}
		\caption{
			(a-b) Temperature dependence of Se 3d and K 3p core levels, respectively. Se 3d peaks are calibrated to the same height and same scaling is applied to K 3p peaks to visualize the changes. (c-d) Temperature dependent ARPES pectra along the $L$-$A$ (119 eV photon energy, k$_z$ = $A$) and $\overline{M}$-$\overline{\Gamma}$ (99 eV photon energy, k$_z$ = 0.17$\Gamma$$A$) directions, respectively. Orange arrows in d mark UB and LB. (e-f) Low temperature ARPES data of pristine (left panel) and S3 (right panel). Pristine sample is cleaved at 40 K sample temperature. FVBs due to the PLD are marked for the pristine sample.
		}
	\end{figure*}

	The chemical environments of pristine and K-doped samples are studied by recording core levels. Se 3d spin orbit split states are located at 53.3 eV (3d$_5/2$) and 54.2 eV (3d$_3/2$) binding energies (Figure 1(a)). K-deposition does not induce a prominent change on their spectral shape and binding energies only a slight decrease in photoemission intensity is observed. For the lowest deposition level, K 3p core level exhibits a relatively broad feature centered around 18.3 eV binding energy (Figure 1(b)). As the deposition increases, K 3p peak gradually shifts to at 17.6 eV. These binding energies are consistent with intercalated K impurities within the van der Waals (vdW) gap \cite{starnberg1997exchange,brauer1998electronic}. The shift to lower binding energies is likely due to the formation of isoelectronic K structures at higher concentrations.

	Figures 1(c) and 1(d) depict ARPES maps taken at room temperature along the $L$-$A$-$L$ directions of the surface Brillouin zone (Fig. 1(e)) for the pristine and K-covered 1T-TiSe$_2$ surfaces, respectively. The conduction band shows a clear evolution with K-deposition. An apparent shift towards higher binding energies is observed as K deposition is increased (Figure 1(f-i)). At highest impurity concentration, conduction band splits into two branches marked with orange arrows Figure 1(i). These states are called upper band (UB) and lower band (LB) now on.

	TiSe$_2$ with the highest K-content (S3) is cooled down to 40 K to monitor the behavior of core levels and electronic structure as a function of sample temperature. The Se 3d peaks are quite robust against temperature variations. In contrast, the spectral intensity of the K 2p level notably increases at low temperatures while becoming sharper (Figure 2(b)). Considering the sampling depth of few TiSe$_2$ layers in ARPES, this can be understood as that K impurities from deeper vdW gaps move toward the nearest one to the sample surface.
	
	Figures 2(c) and 2(d) display the evolution of conduction bands in S3 as the sample temperature is reduced. In the data taken with 119 eV photons at 100 K and below, LB shows a sharp parabolic dispersion at the $L$-point while UB is no longer visible in the spectra (Figure 2(c)). In contrast, ARPES data taken with a photon energy of 99 eV at the same temperature host LB and UB simultaneously. Both exhibit a parabolic dispersion with an increased energy splitting of 130 meV (Figure 2(d)).
	
	The possibility of imperfections causing the observed band splitting is also ruled out by examining the ARPES maps at different spots on the sample surface (see Figure S2 in Ref. \cite{SI}). The same splitting is consistently observed along with identical core levels. However, in regions with lower K concentration, the splitting is much less. This further confirms the direct correlation between the intercalated K-impurities and the observed band splitting. Furthermore, we corroborate our observations by performing the same experiment on a separately cleaved sample (see Figure S3 in Ref. \cite{SI}). The data are reproducible, as the conduction band consistently shows the same splitting upon cooling the K-deposited sample.
	
	The low-temperature electronic structure of K-deposited sample is compared with that of the pristine one (Figures 2(e-f)). FVBs resulting from the 2a$\times$2a$\times$2c superstructure periodicity are clearly resolved in the surface electronic structure of the pristine sample (left panel in Figures 2(e-f)) (also see Figure S4 in Ref. \cite{SI}). In contrast, K-doping completely suppresses the PLD, as evidenced by absence of the FVBs (right panel in Figures 2(e-f)). For the further comparison, temperature evolution of the band structure in pristine TiSe$_2$ is also given in Figure S5 of \cite{SI} where folding of the conduction and valence bands are shown.

	Splitting of the conduction band into two branches, correlated with increasing impurity concentration near the surface region strongly suggests the detachment of top TiSe$_2$ layer from rest of the material. This can be verified through photon energy-dependent ARPES data, which enables the distinction between 2D and 3D dispersion. ARPES data for several photon energies are presented in Figure 3(a). It is evident that the upper band (UB) disperses as a function of photon energy, while the lower band (LB) remains unchanged in terms of its binding energy and dispersion. This is further illustrated in Figure 2(b), which depicts the in-plane versus out-of-plane momentum at the Fermi level. LB shows a non-dispersive feature along the k$_z$-direction (marked with dashed yellow lines in Figure 3(b)), while the bulk bands exhibit 3D character (marked with dashed red lines in Figure 3(b)). Photon energy dependence of the valence band in the zone center is also discussed in Figure S5 of \cite{SI}. in which a quasi-two dimensional band dispersion is weakly resolved. Furthermore, the Fermi surface at k$_z$ = $\Gamma$ features two ellipsoidal pockets at the $M$ points (Figure 3(c)). The inner pocket is smeared out at the $L$-points, while the outer pocket remains unchanged (Figure 3(d)), further confirming the distinct dispersion of the two bands.

	\begin{figure}[t]
		\centering
		\includegraphics[width=8cm,height=7.197cm]{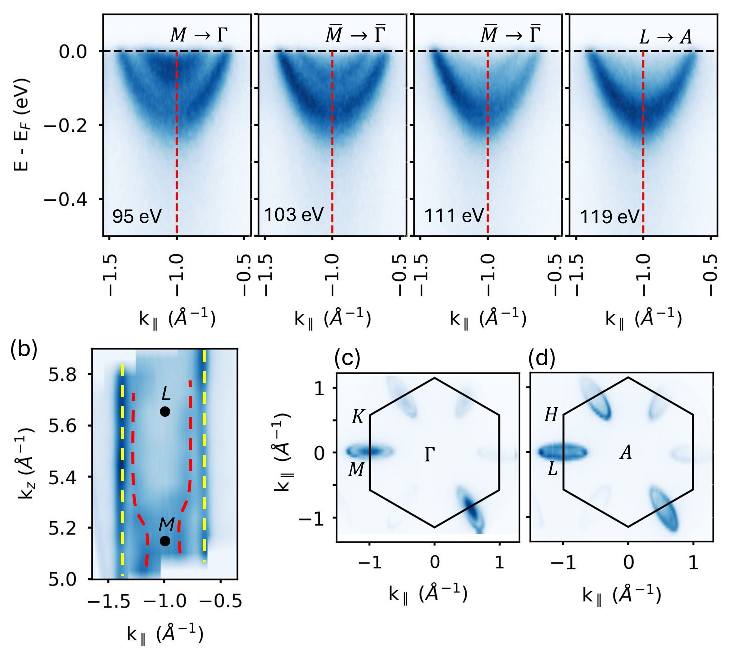}
		\caption{
			(a) Photon energy dependent ARPES maps taken from S3 along the $\overline{M}$ - $\overline{\Gamma}$. (b) k$_z$ versus k$_\parallel$ dispersion at the Fermi level. Dashed yellow and red lines represent the out of plane dispersion of LB and UB, respectively. (c-d) Fermi surfaces at k$_z$ = $\Gamma$ and $L$-points. Black hexagons in c and d represent the 2D Brillouin zone. Sample temperature was kept at 40 K during the measurements.
		}
	\end{figure}

	\begin{figure}[t]
		\centering
		\includegraphics[width=7.5cm,height=7.840cm]{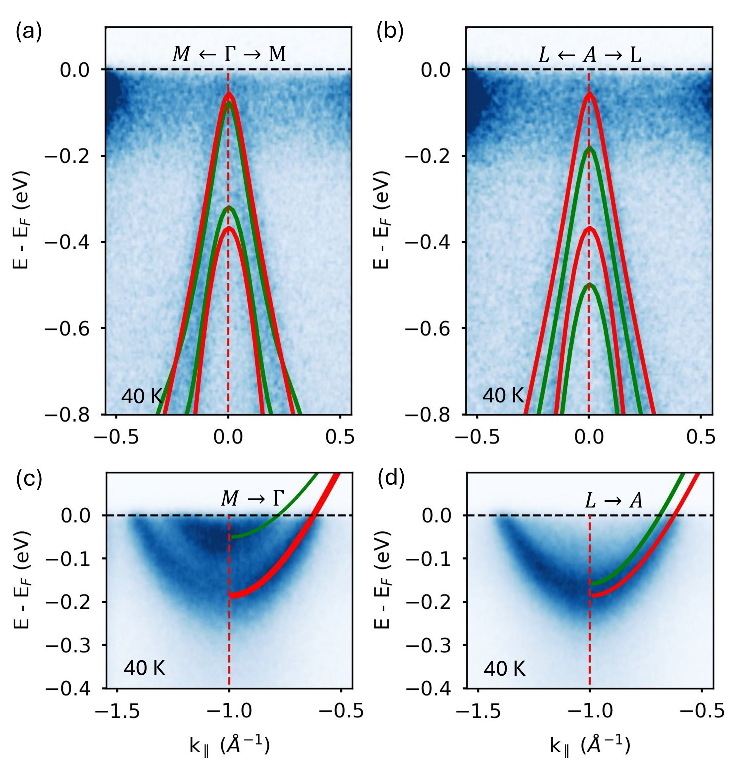}
		\caption{
			(a-b) Experimental valence band spectra along the $M$ - $\Gamma$ - $M$ and $L$ - $A$ - $L$, directions, respectively. (c-d) Experimental conduction bands along the $M$ - $\Gamma$ and $L$ - $A$ directions, respectively. Calculated band structures of monolayer (red lines) and bulk (green lines) TiSe$_2$ are superimposed onto the measured band structures. Color contrast in a and b is enhanced to increase the visibility of valence bands. Experimental data are collected from S3 at 40 K sample temperature and spin orbit coupling is included in the band structure calculations.
		}
	\end{figure}

	Experimentally, LB and UB, respectively, originate from the quasi-freestanding TiSe$_2$ layer and bulk TiSe$_2$ with distinct electronic dispersion. Therefore, the theoretical band structure of each system can be compared with the low temperature experimental band structure of S3. We calculate the band structure by incorporating the Hubbard U term (see Figure S6 in Ref. \cite{SI}). This is shown to be a better approach to describe experimental band structure of the normal phase in TiSe$_2$ \cite{bianco2015electronic,hellgren2021electronic}. Furthermore, we use the chemical potential of bulk TiSe$_2$ since the electron doping from K-impurities are negligible. We do not observe significant changes in the valence band (Figure 4(a-b)), and this is anticipated since K-doping primarily affects the Ti 3d atomic orbitals \cite{jeong2024dichotomy}. This selective modification suggests that the electronic structure near the Fermi level remains largely intact, even as the doping influences lower-energy states associated with the Ti orbitals. However, the dispersion characteristics of LB and UB conduction bands at the $M$ and $L$-points  are well captured by the computed band structures for monolayer and bulk samples, respectively (Figures 4(c-d)). This strong correlation between the two datasets provides robust evidence for the successful formation of a quasi-freestanding TiSe$_2$ layer through K-intercalation.

	The formation of a quasi-freestanding layer of TiSe$_2$ indicates a significant expansion in the vdW gap. Considering a sampling depth of 10 \r{A} in ARPES and the simultaneous observation of 2D and 3D bands near the sample surface strongly suggests that K impurities are predominantly located within the first vdW gap, just beneath the surface. Otherwise, multiple 2D bands originating from different layers would be expected in the ARPES data. Hence, the elemental stoichiometry near the sample surface can be calculated using the area of core-level peaks and photoionization cross-sections \cite{yeh1985atomic}. We calculate a composition of K$_{0.26}$TiSe$_2$ using Se 3d and K 2p core levels (see Figure S7 in Ref. \cite{SI}).
	
	A nearly identical compound can be modeled to study the crystal and electronic structure in more detail (Figure 5(a-b)). Here, we use K$_{0.25}$TiSe$_2$ as a prototype and compute the lattice constants and electronic dispersion, particularly along the k$_z$ direction. The in-plane lattice parameters remain almost unchanged upon the inclusion of K impurities in the vdW gap, while the c-lattice parameter increases from 5.99 \r{A} to 7.92 \r{A} (Figure 5(a-b)). The corresponding band structures are compared with those of pristine TiSe$_2$ in Figure 5(c-d). The lattice expansion along the c-axis induces a crossover from 3D to 2D dispersion, as observed by the negligible band dispersion along the $\Gamma$-$A$ and $M$-$L$ directions. We also evaluate the lattice parameters for various K amounts. The c-lattice parameter initially shows a prominent increase and then starts shrinking (see Figure S8 in Ref. \cite{SI}). Very similar results have also been reported for K-intercalated TiS$_2$\cite{zhang2022k+}. Such behavior is probably attributed to formation of a strong chemical bonding between K and Se atoms at higher doping levels.

	\begin{figure}[t]
		\centering
		\includegraphics[width=8cm,height=8.856cm]{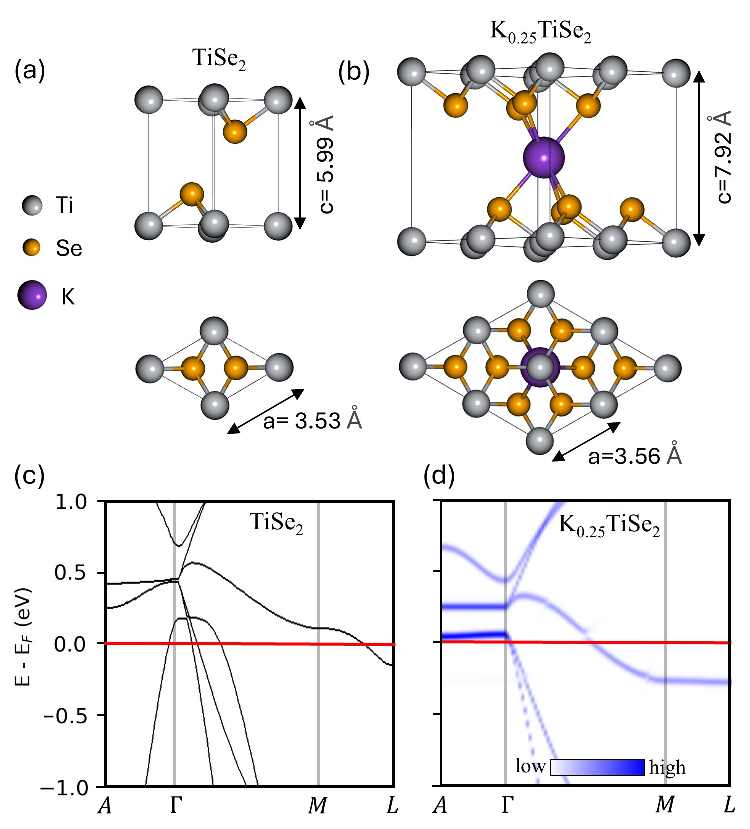}
		\caption{
			(a-b) Ball and stick representations of pristine TiSe$_2$ and K$_{0.25}$TiSe$_2$ supercell, respectively. Upper and lower panels display side and top views. (c-d) Electronic structures of pristine TiSe$_2$ and K$_{0.25}$TiSe$_2$ supercell along the $A$-$\Gamma$-$M$-$L$ direction. Supercell band structure is unfolded into the Brillouin Zone of TiSe$_2$ unit cell. In these band structure calculations, spin orbit coupling and U-term are not included.
		}
	\end{figure}

	Confinement of the electronic structure near the surface region might also be considered as a candidate for band splitting. However, the distinct dispersion characteristics of the two bands along the k$_z$ direction and the absence of significant electron doping rule out this scenario. Recently, quantum well states (QWSs) have been induced in TiSe$_2$ upon deposition of K-impurities on the sample surface at low temperature \cite{jaouen2023carrier}. These impurities act as ad-atoms, randomly distributed on the surface, and doped the system with electrons. This leads to the emergence of QWSs, but with notably smaller splitting energy than what is observed here between the conduction bands of quasi-freestanding layer and bulk part of the material.
	
	Furthermore, weakening of the spectral intensity in valence bands is also ambiguous, while the Ti 3d-derived conduction band exhibits strong photoemission intensity even after K-deposition. Additionally, the quasi-freestanding layer is a semimetal with an overlap between the conduction and valence bands. The combined effect of the modified band gap and weaker spectral weight of the valence band can refer to weak p-d interaction, causing suppressed PLD from the ARPES perspective \cite{rossnagel2010suppression, jeong2010electronic}. These effects are not observed in bulk samples after K deposition at low temperatures or in monolayer TiSe$_2$ samples grown on a substrate \cite{jaouen2023carrier, chen2015charge}. Both show an indirect bulk band gap and strong spectral weight in the valence band region. Therefore, in contrast to our observations, PLD survives in these samples.
	
	Upon room temperature K-deposition, ARPES electronic structure suggests the formation of quasi-freestanding TiSe$_2$ layer while PLD is suppressed. This observation is consistent with the evolution of electronic structure in monolayer TiSe$_2$ upon deposition of K-impurities \cite{kolekar2018controlling}. Similar to the our case, FVBs disappears and the energy gap between conduction and valance bands closes. Furthermore, a scanning tunneling microscopy study in K-deposited bulk TiSe$_2$ also revealed the intercalation of impurities and suppression of PLD followed by the shrinking in bulk band gap, attributed a crossover from lattice distortion to superconductivity \cite{zhang2018unveiling}. Furthermore, similar to our case, intercalation of impurities also induces free-standing MoS$_2$ monolayer with a smaller band gap compared to its bulk form \cite{eknapakul2014electronic}.
	
	In summary, we demonstrate an ARPES study showing the formation of a quasi-freestanding TiSe$_2$ layer induced by room-temperature deposition of K-impurities, followed by sample cooling. This intercalation method could be an effective way to control sample dimensionality and, consequently, the electronic properties. Furthermore, absence of PLD upon K-disposition can lead to emergence of other physics in this material. Therefore, our result can guide future studies on phenomena competing with periodic lattice distortions, such as superconductivity.

	\section{ACKNOWLEDGMENTS}
	
	This research used resources ESM (21ID-I) beamline of the National Synchrotron Light Source II, a U.S. Department of Energy (DOE) Office of Science User Facility operated for the DOE Office of Science by Brookhaven National Laboratory under Contract No. DE-SC0012704. We have no conflict of interest, financial or other to declare. This research also used the research grant from Xiamen University Malaysia (Grant No. IORI/0007).

\end{document}